# Topological phases and nonreciprocal edge states in non-Hermitian Floquet Insulators

Mengyao Li[1,2], Xiang Ni[1,2], Matthew Weiner[1,2], Andrea Alù[3,2,1], and Alexander B. Khanikaev[1,2]†

[1]Department of Electrical Engineering, Grove School of Engineering, City College of the City University of New York, 140th Street and Convent Avenue, New York, NY 10031, USA
[2]Physics Program, Graduate Center of the City University of New York, New York, NY 10016, USA
[3]Photonics Initiative, Advanced Science Research Center, City University of New York, New York, NY 10031, USA

†Corresponding author. Email: khanikaev@gmail.com

**Topological phases in quantum and classical systems have been of significant recent interest due to their fascinating physical properties. While a range of different mechanisms to induce topological order have been introduced, a quest for the most viable solution for practical systems is still open. Floquet topological insulator represent one of possible venues to engineer topological phases, yet they have been so far largely restricted to temporal modulation of Hermitian potentials. On the other hand, in many physical systems, including acoustic and optical systems, modulating loss or gain can be more straightforwardly achieved. Two of such examples are graphene, which enables strong modulation of its conductivity due to saturable absorption, and quantum wells where population inversion can be achieved in an ultrafast manner. On the other hand, non-Hermitian Floquet potentials have not been shown to yield novel topological phases to date. It is therefore of great interest to explore time-modulated non-Hermitian potentials in periodic lattices, and the emergence of topological phases associated with them. Here we demonstrate that non-Hermitian Hamiltonians can indeed result in topological phases supporting nonreciprocal edge states propagating without dissipation, as well as new regimes of dissipative and amplifying topological edge transport.**

The discovery of novel topological phases of quantum matter have triggered active research in a broad range of classical systems, from acoustics and mechanics to photonics [1, 2, 3, 4, 5, 6, 7, 8, 9, 10, 11, 12, 13, 14, 15, 16, 17, 18, 19, 20, 21, 22, 23]. Characterized by the presence of robust edge states, their classical counterparts open the opportunity of unusual signal transport and wave manipulation in optical and mechanical metamaterials. In this regard, designer topological materials hold a special promise for new ways of transmitting, manipulating and processing information. However, due to their very nature, no classical topological phases can be induced by time-reversal symmetry alone. Two common approaches to overcome this limitation have been explored: breaking time-reversal symmetry or exploiting symmetry protected phases relying on underlying spatial symmetries.

Unfortunately, the means of breaking TR symmetry for mechanical or optical waves are quite limited, and are often hard to implement in practice due to charge neutrality of phonons and photons, leading to their weak interaction with magnetic fields, and weak magneto-optical and magneto-elastic effects. Symmetry protected topological phases, on the other hand, have obvious limitations imposed by the restricted nature of their robustness, which is vulnerable to any symmetry violating perturbations, as they are bound to obey reciprocity. For this reason, there is currently an ongoing pursuit in establishing classical potentials capable of inducing topological nontrivial phases in practical technological platforms. In this respect, Floquet topological phases in classical system with periodically varying potentials can be considered as a viable alternative to the approaches relying on symmetries [6, 7, 9, 19, 24, 25, 26, 27, 28, 29, 30, 31, 32, 33]. Moreover, Floquet systems can show new rich physical properties, for example, in addition to topological Chern-class phases, Floquet systems have been reported to host another unique topological phase, referred to as anomalous [31, 34, 35].

Besides aspects related to topological properties, due to the urgent need in nonreciprocal devices for photonic and acoustic applications, there is a significant interest in utilizing time-modulation to achieve nonreciprocal propagation [6, 9, 33]. For this reason, combining nonreciprocity with topological robustness may open exciting opportunities for practical technology. Indeed, it has already been shown that such an approach allows to achieve high performance and broadband nonreciprocal isolators and one-way leaky antennas in Hermitian acoustic Floquet systems [33].

Unfortunately, temporal modulation itself in either optical or mechanical systems is a rather nontrivial task, especially at high frequencies. This is particularly problematic in photonics, due to the extremely weak character of electro-optical and nonlinear phenomena, which could be used to modulate dielectric permittivity or high-frequency conductivity of materials. Interestingly, this limitation does not necessarily apply in such strong terms to the imaginary part of the dielectric constant [36], as both gain and loss can be modulated with relatively large amplitude, e.g., in systems with saturable absorption such as graphene and reduced graphene oxide, as well as in optically active media, such as quantum wells and quantum dots, where one can achieve reversal of loss to gain by increasing the amplitude of the modulating pump signal. Also the modulation of Drude conductivity can be achieved by electron-hole plasma generation using ultrafast optical pumping. Fast relaxation time in these system may further enable modulation at rates high enough to yield topological Floquet phases in infrared and terahertz domains, provided that modulation of dissipative or amplifying responses in time yields topological properties.

Gain and loss modulation thus could be exploited to induce topological responses, although it is not at all obvious that such modulation may yield topologically nontrivial phases. Indeed, it has previously been suggested that static non-Hermitian potentials cannot induce topological phases [37], and it may be tempting to extend this conclusion to time-dependent potentials. However, the role of non-Hermitian corrections to Hermitian topological Hamiltonians has been recently explored, and it was shown that some interesting phenomena, including exceptional points in bulk and edge topological spectra, and topological transitions induced by gain and loss, have been demonstrated [38, 39, 40, 41, 42, 43, 44, 45, 46, 47, 48].

In order to understand whether topological phases are feasible in systems with time-modulated gain and loss, here we study the effect of time-periodic non-Hermitian potentials on topologically trivial Hermitian systems. We demonstrate that, in the case of time-driven non-Hermiticity, time-modulation can lead to topologically nontrivial Floquet phases of Chern and anomalous type. Moreover, we find that topological edge states in such systems can be rendered dissipationless by effectively averaging gain and loss over the modulation period (without relying on parity-time, or other symmetries). Alternatively, non-Hermitian Floquet systems can be driven into a regime of purely amplifying or dissipating edge transport, which can be important for applications, such as in topological lasers [49, 50, 51].

Before proceeding to our numerical results demonstrating the outlined topological regimes, we start with the analytical theory supporting such responses. We consider the topologically trivial time-independent Hermitian Hamiltonian $\widehat{\mathcal{H}}_0$, whose eigenstates satisfy the time-dependent Schrodinger equation $\widehat{\mathcal{H}}_0|\psi_0\rangle = i\partial_t|\psi_0\rangle$, and its temporal evolution described by the unitary operator $\widehat{U}_0(t) = \exp[-i\widehat{\mathcal{H}}_0 t]$. The time-periodic non-Hermitian perturbation $\widehat{\mathcal{V}}(t) = i\widehat{V}_S(t)$, where $\widehat{V}_S(t) = \widehat{V}_S(t+T)$ is the Hermitian time-periodic operator, added to $\widehat{\mathcal{H}}_0$ to describe modulation of gain and loss. The periodic character of the perturbation implies that the standard stroboscopic evolution approach can be utilized to describe the system dynamics. The trivial dynamics of the system (described by $\widehat{U}_0(t)$) can be conveniently eliminated in the interaction representation picture, in which the Schrodinger equation assumes the form $i\widehat{V}_I(t)|\psi_I\rangle = i\partial_t|\psi_I\rangle$, where $\widehat{V}_I(t) = \widehat{U}_0^{-1}(t)\widehat{V}_S(t)\widehat{U}_0(t)$ is a periodic non-Hermitian operator in the interaction representation. Due to the unitary property of the evolution operator $\widehat{U}_0(t)$, the operator $\widehat{V}_I(t)$ is Hermitian. The evolution of the perturbed system is then described by the evolution operator

$$\widehat{U}_I(t_1, t_2) = \widehat{T}\exp[\int_{t_1}^{t_2}\widehat{V}_I(t)dt], \qquad (1)$$

which represents a time ordered product of non-unitary operators (exponents of non-Hermitian operators) and therefore in general can be non-unitary. Stroboscopic evolution of the system allows us to describe the effects of gain-loss modulation in terms of an effective perturbing potential $\widehat{V}_{eff} = \frac{1}{i}\log[\widehat{U}_I(0,T)]$, in which case the system can be characterized by an effective Hamiltonian of the form $\widehat{\mathcal{H}}_{eff} = \widehat{\mathcal{H}}_0 + \widehat{V}_{eff}$ [26].

Interestingly, despite the fact that neither $\widehat{\mathcal{H}}_0$ nor the instantaneous perturbing potential $\widehat{\mathcal{V}}(t_0 = \text{const})$ may yield topological phase, the resultant time-modulated system described by the effective Hamiltonian $\widehat{\mathcal{H}}_{eff}$ can in fact be topological. For this statement to be correct, the effective potential $\widehat{V}_{eff}$ should contain a (topologically non-trivial) Hermitian part, which implies that the stroboscopic evolution operator $\widehat{U}_I(0,T)$ should not be purely anti-unitary. This is indeed possible, due to the fact that Hermitian operators do not form closed commutative algebra; therefore, in general, the product of exponents of Hermitian operators $\widehat{V}_I(t)$ in Eq. 1 may give rise to an effective stroboscopic potential $\widehat{V}_{eff}$ containing both Hermitian and anti-Hermitian parts. Surprisingly, as we show below, the anti-Hermitian part may have vanishing eigenvalues (for some distributions of gain and loss modulation), thus making the system dissipationless. This has some analogy to PT-symmetric systems [48], but applies to a broader class of non-Hermitian topological systems.

To further show that time-modulated gain and loss may render a topologically nontrivial effective Hamiltonian, we first consider the geometry illustrated in Fig. 1. It consists of a Kagome lattice with unit cell containing three identical single-mode resonators with resonant frequency $\omega_0$. Assuming intra-cell and inter-cell hopping amplitudes, $\kappa$ and $j$, respectively, in the absence of time-modulation we obtain the unperturbed Hamiltonian

$$\hat{\mathcal{H}}_0 = \begin{pmatrix} \omega_0 & \kappa + je^{i(\frac{1}{2}k_x+\frac{\sqrt{3}}{2}k_y)} & \kappa + je^{-i(\frac{1}{2}k_x-\frac{\sqrt{3}}{2}k_y)} \\ \kappa + je^{-i(\frac{1}{2}k_x+\frac{\sqrt{3}}{2}k_y)} & \omega_0 & \kappa + je^{-ik_x} \\ \kappa + je^{i(\frac{1}{2}k_x-\frac{\sqrt{3}}{2}k_y)} & \kappa + je^{ik_x} & \omega_0 \end{pmatrix}, \quad (2)$$

where $\mathbf{k} = (k_x, k_y)$ is the two-dimensional Bloch vector. The spectrum of $\hat{\mathcal{H}}_0$ is shown in Fig. 1b and it reveals three bands, corresponding to one monopolar and two dipolar states. The triangular symmetry of the lattice yields two Dirac points, due to degeneracies between dipolar and monopolar modes at the K and K′ points at the corners of the hexagonal Brillouin zone (BZ). In addition, due to rotational and time-reversal symmetries, the system possesses a degeneracy between dipolar modes at the Γ point of BZ.

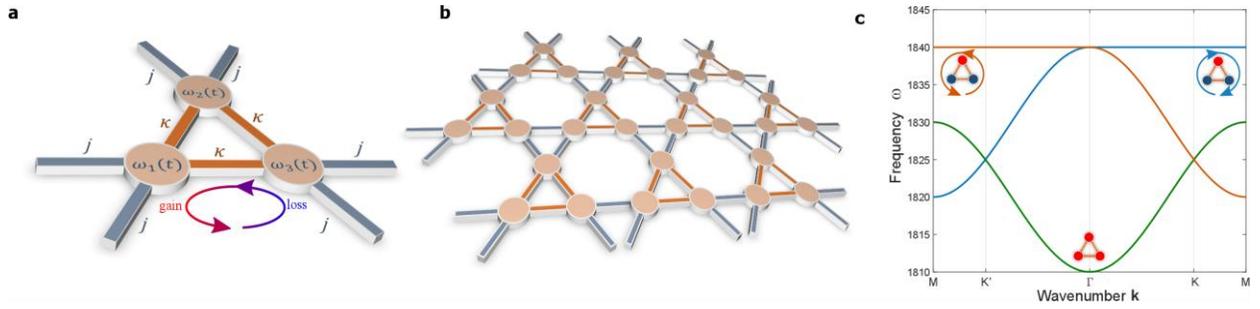

**Fig. 1| Non-Hermitian Floquet Kagome lattice. a**, **b** unit cell and periodic arrangement of time-modulated Kagome lattice with on-site frequency (energy) $\omega_n(t) = \omega_0 + iv_n(t)$, and inter-cell and intra-cell coupling $j$ and $\kappa$. Modulation of gain and loss $v_n(t)$ follows the rotational pattern shown in **a**. **c** Band structure and modal profiles of the states supported by the unmodulated Kagome lattice, revealing degeneracies at K, K′, and Γ points.

Non-Hermitian periodic modulation is introduced by adding a piecewise time-dependent potential $\hat{V}(t) = i\hat{V}_S(t)$, where $\hat{V}_S(t)$ is a diagonal matrix with diagonal elements $\text{diag}(\hat{V}_{S1}) = v(1, -\frac{1}{2}, -\frac{1}{2})$ for the first one-third of the Floquet period ($t \in (0, T/3]$), $\text{diag}(\hat{V}_{S2}) = v(-\frac{1}{2}, 1, -\frac{1}{2})$ for the second one-third of the period ($t \in (T/3, 2T/3]$), and $\text{diag}(\hat{V}_{S3}) = v(-\frac{1}{2}, -\frac{1}{2}, 1)$ for the last one-third of the period ($t \in (2T/3, T]$), i.e., the on-site loss/gain in the three resonators is modulated with a phase delay of 120 degrees (Supplement S2), and parameter $v$ is the depth of the non-Hermitian modulation. Similar modulation protocol of Hermitian modulation have been previously shown to yield nonreciprocal and topological effects [33, 52].

For illustrative purposes, here we limit the analytical treatment and effective Hamiltonian description to the proximity of the degeneracy between dipolar modes at the Γ point. In this case, the two-band approximation can be used, and the model yields a simple analytical result. The resultant effective 2x2 Hamiltonian acts on circularly polarized dipolar modes, and, up to the second-order in wavenumber, it assumes the form

$$\widehat{H}'_{eff} = \frac{j(k_x^2+k_y^2)}{4}\hat{\sigma}_0 + \frac{j(k_y^2-k_x^2)}{4}\hat{\sigma}_x + \frac{jk_xk_y}{2}\hat{\sigma}_y + \frac{\sqrt{3}}{8}v^2\hat{\sigma}_z,$$

from which the $\hat{\sigma}_z$ term can be interpreted as an effective magnetic bias opening a topological bandgap between dipolar bands at the $\Gamma$ point. Importantly, the effective potential $\frac{\sqrt{3}}{8}v^2$ is a real number, despite the fact that the modulation applied to the system is purely imaginary. As we confirm below by numerical calculations, this conclusion holds beyond our approximations, and regimes exist when both bulk and edge states have purely real spectra.

These analytical results are validated with numerical simulations, in which we assume a continuous periodic time-dependent potential $\widehat{\mathcal{V}}(t) = i\widehat{V}_S(t)$, where $\widehat{V}_S(t)$ has a harmonic form. As before, the on-site gain/loss is modulated with a phase shift of 120 degrees between resonators $\text{diag}(\widehat{V}_S(t)) = v\left[\sin(\omega t), \sin(\omega t + \frac{2\pi}{3}), \sin(\omega t + \frac{4\pi}{3})\right]$, and the unperturbed Hamiltonian $\widehat{\mathcal{H}}_0$ in Eq. (2) is unchanged.

The effective Hamiltonian is calculated by numerically evaluating the product of matrix exponents at discrete times, with 1800 steps per modulation period, ensuring excellent convergence. The band structure obtained for the effective Hamiltonian for different modulation depths are shown in Fig. 2. We clearly see that, in the case of weak modulation $v = 0.09\omega_0$, the band structure is primarily affected near the points of former degeneracies, at K/K′ and $\Gamma$ points of BZ, respectively, where complete bandgaps are open by the modulation. Inspection of the complex band structure in Fig. 2b shows that the bands retain their purely real character, despite the presence of gain and loss. Note that this is not due to PT symmetry, but rather due to averaging of gain and loss over one modulation period, which, for this specific choice of modulation protocol, appears to balance off the effects of gain and loss on the modes for all wavenumbers, leading to a pseudo-Hermitian real-valued spectrum.

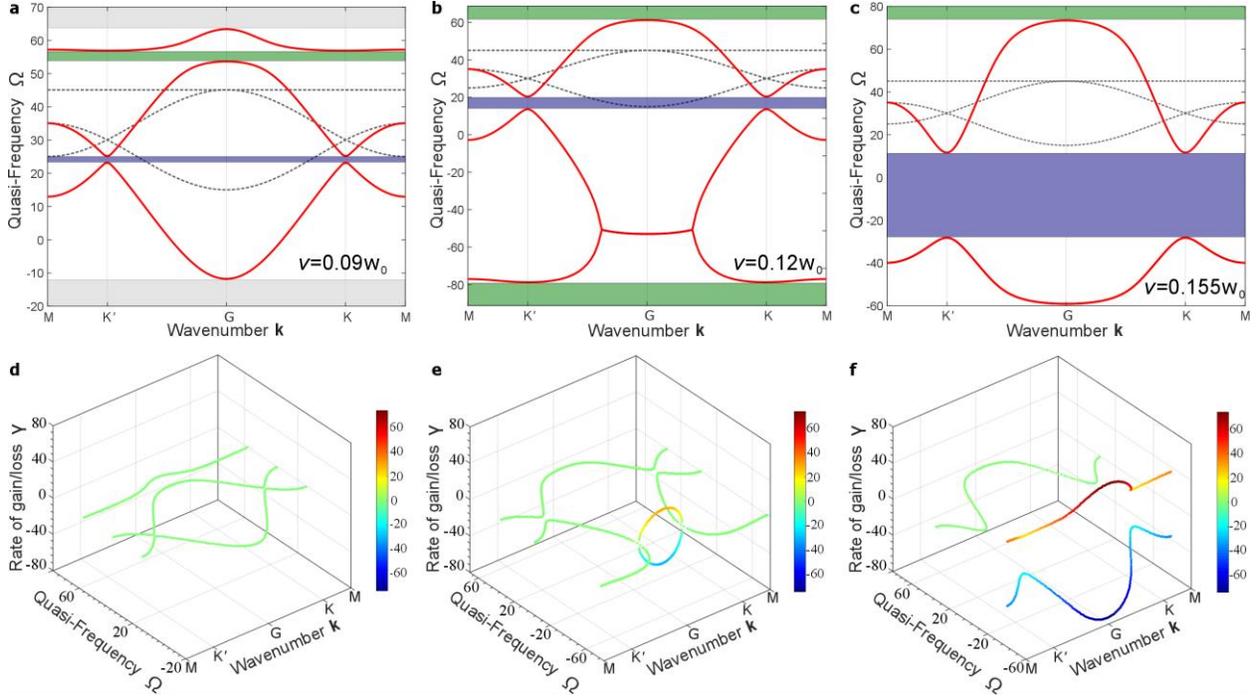

**Fig. 2| Complex photonic band structure of non-Hermitian Floquet Kagome lattice. a, b, c** Real part of eigenvalues of the quasi-frequency (quasi-energy) of time-modulated gain and loss in the structure, with progressively increasing modulation depth $v$. **d, e, f** Complex eigenvalues of quasi-frequency of cases shown in **a**, **b**, and **c**, respectively.

As seen from Figs. 2c,d and 2e,f, the pseudo-Hermitian regime does not hold for larger values of modulation depth, and some dramatic changes appear in the band structure. As the modulation depth is increased, we first observe increased separation between higher frequency dipolar bands, and eventual collision of one of the bands with lower lying monopolar band when modulation depth is $v = 0.12\omega_0$. In the quasi-frequency description, this is due to the high frequency band entering the diagram from the low-frequency side. The collision of bands leads to degeneracy in the real quasi-frequency, with degeneracy being lifted in the imaginary part of the spectrum. Therefore, the wavenumbers exist where the spectrum experiences transitions from real to complex valued, which represent exceptional points of the Floquet spectrum. Interestingly, the closure of the band gap separation of former high frequency and low frequency bands does not affect bandgaps open by the modulation at the K and K' points. In addition, one of the former dipolar bands remains purely real-valued, even for increased modulation strength.

Further increase in the modulation depth ($v = 0.155\omega_0$) leads to even more nontrivial changes in the spectrum. In particular, the exceptional points gradually move towards the edges of the BZ, until the degeneracy in imaginary quasi-frequency is completely removed (lastly at the K/K'-points). At this point, the real spectrum is completely degenerate, and the bandgap exists only in the imaginary quasi-frequency direction. This regime resembles the anomalous Floquet regime of Hermitian Floquet systems, since the gap appears between bands of different Floquet orders, with the difference that in our case the gap appears in the imaginary and not the real part of the spectrum. This raises the question of whether such a transition, accompanied by gap opening in the imaginary plane, leads to topological features, and to the emergence of topologically protected edge states.

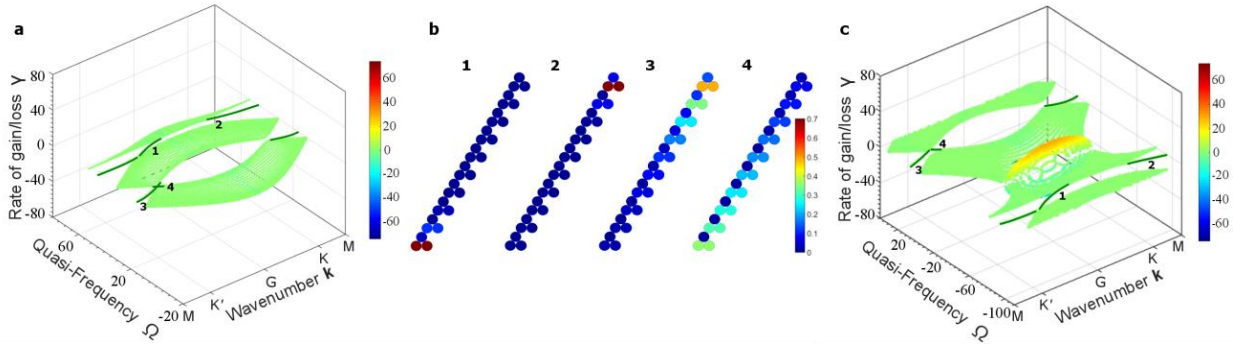

**Fig. 3| Complex photonic band structure of non-Hermitian Floquet Kagome supercell for weak modulation. a**, **c** Complex eigenvalues of quasi-frequency (quasi-energy) of the time-modulated structure with modulation depth $v = 0.09\omega_0$ and $v = 0.12\omega_0$ respectively. The edge modes of both geometries are shown in **b** (note that the supercell was truncated from 20 to 10 to highlight better the mode profile).

Emergence of band-crossing edge states is one of the main signatures of topologically nontrivial regimes. In order to see whether such states emerge within the bandgaps (both in real and imaginary frequency directions), we calculated the band structure of a supercell consisting of 1x20 trimers (unit cells) of modulated crystals terminated on upper (bearded-like) and lower (straight) edges, and with periodic boundary conditions imposed in the horizontal direction. The resultant complex band structure for the cases of weak and intermediate modulation is shown in Fig. 3, and it clearly reveals a set of new states within the bulk band gaps. The wave-functions of these states appear to be localized at the edges of the system (Fig. 3 insets), and therefore represent edge states induced by the gain-loss modulation. Their dispersion is nonreciprocal, due to the selected rotating modulation scheme, and one-way transport along the upper (lower) edge takes place in the positive (negative) direction. Just as for Hermitian Floquet systems, the propagation direction reverses when the rotation direction is flipped [33]. Note that the edge spectrum for two cuts is not symmetric, and the respective bands do not transform one into another under $k_\parallel \to -k_\parallel$ transformation. This asymmetry in edge spectrum is related to the fact that the upper (bearded) and lower (straight) cuts are not equivalent.

For the case of weak modulation, the edge states appear to be purely real, thus indicating that gain and loss are compensated on average over a single period for the given cuts. Note that a different modulation protocol, in particular a different modulation phase, may correspond to edge states with a small imaginary quasi-frequency component. The real bulk spectrum also allows us to immediately apply the standard approach of calculating Berry curvature and Chern numbers for the bands, which are found to be $C = (1, -2, 1)$ for three bands counted from lowest eigenfrequency up in Figs 2a and 3a. In accordance with the bulk-boundary correspondence principle, these numbers agree well with the number of edge states, and with the direction of the modes on a particular cut, thus further confirming that non-Hermitian time-modulated potentials can yield effective Hermitian (pseudo-Hermitian) topological phases in the stroboscopic picture.

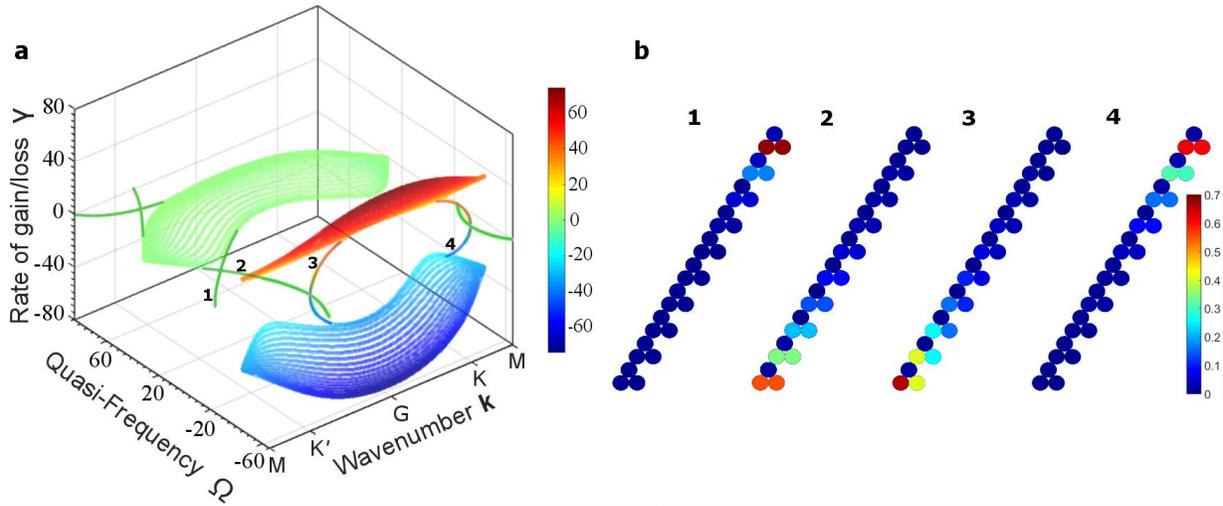

**Fig. 4| Complex photonic band structure of non-Hermitian Floquet Kagome Supercell for strong modulation. a**, **c** Complex eigenvalues of quasi- (quasi-energy) of the time-modulated gain and loss in the structure with modulation depth $v = 0.155\omega_0$. The typical edge modes of both cases are shown in **b**. **3** and **4** represent the two branches of the complex valued edge band.

With an increase in modulation depth, the topological character of the edge states ensures that they will prevail, as long as the gaps remain open. This is confirmed by our calculation for the "intermediate" modulation depth $v = 0.12\omega_0$ shown in Fig. 3c, where, despite closure of a trivial gap between different Floquet orders (between lowest and highest bands), and emergence of exceptional points and complex-valued bulk spectrum, the edge spectrum remains purely real-valued. Note that, although for intermediate modulation we have effectively entered the anomalous Floquet regime, due to the crossing of bulk bands belonging to different Floquet orders, this crossing does not lead to a new topological phase, as no new gaps arise in either the real or imaginary parts of the spectrum.

This picture is dramatically modified if we further increase the modulation depth and enter the regime of "strong" modulation ($v = 0.155\omega_0$), which is characterized by opening of a complete band gap between the first and third bands of different Floquet orders in the imaginary quasi-frequency dimension. As the bands appear to be spectrally separable again, this allows us to calculate Chern numbers, which appear to be $C = 1$ for complex valued bulk bands and $C = -2$ for real valued band. This agrees well with the previous calculation in the pseudo-Hermitian regime; indeed, the real-valued band is still separated from the other two, and its topological invariant is therefore preserved. The other two bands merge together and split again, now in the imaginary frequency direction, but, under this transition, they again acquire the same values of topological invariant, which can be explained from the symmetry of the spectrum. Indeed, the sum of Chern numbers of all three bands vanish, leaving us with the total Chern number of the two complex bands, equal to 2. As the complex bands are clearly symmetric, i.e., they have identical real part of the spectrum, and complex conjugate imaginary part, they are poised to have identical Chern numbers. This heuristic argument is confirmed by a direct inspection of the wave-functions in the complex bulk bands, which appear to be identical up to a similarity transformation (inversion in the direction parallel to the edge).

The above conclusions about the Chern numbers directly translate into the properties of the edge spectrum in the strongly modulated non-Hermitian case. However, the complex spectrum contains an important difference from the case of Hermitian (and pseudo-Hermitian) systems, which should affect the way the edge and bulk states interconnect in both real and imaginary parts of the spectrum. Thus, according to the bulk-boundary correspondence, we should observe two edge bands each interconnecting one of the complex bulk bands with the real-valued bulk band. We indeed see that the edge bands interconnect the bulk bands, but this connectivity takes place via a new set of states within the complex spectrum that interconnect the two complex bulk bands with each other. These new states are not found in the bulk spectrum calculated for an infinite crystal (Fig. 2f), thus implying that they are related to the presence of the edges. Indeed, an inspection of the wave-function of these states shows that they are localized to the edges (Fig.4b). We therefore conclude that the connectivity of the edge and bulk spectra takes place via exceptional points in the edge spectrum. The main consequence of this observation is that the edge spectrum of the same system can be either real or complex valued.

As a result, the edges of the system can support either (i) a conventional lossless (and gainless) topologically robust edge transport via edge states with real spectrum, (ii) topologically robust propagation exponentially attenuating in time, and, finally, the most intriguing regime (iii) topologically robust propagation that amplifies exponentially in time. The latter regime can be of importance for practical applications, in particular, for designing topologically robust active optical devices, including topological lasers [49, 50, 51].

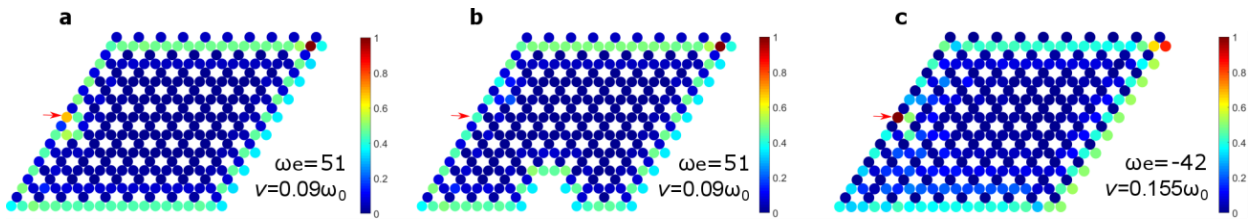

**Fig. 5|Edge states of finite Kagome structure.** We use a source at one site on the boundary (red arrow) to excite the edge states. $\omega_e$ is the excitation quasi-frequency of the source. **a** Edge state induced by weak modulation, with excitation quasi-frequency at the edge between two doublet bands. **b** Same as **a**, but with a defect at the edge, highlighting the robustness of the edge state. **c** Edge state for strong modulation. We picked an excitation quasi-frequency corresponding to the real part of the complex valued edge states (and observe exponential growth in time).

To further understand the behavior of the edge states, we performed modelling of large domains of crystals, shown in Fig. 5, with edge states excited by a point source located in the proximity of one of the edges (indicated by arrow in Fig.5). As expected, only edge modes within the quasi-frequency range of the topological bandgap are excited, and we observe their reflectionless propagation across sharp corners of different cuts, and around deliberately introduced defects.

Note that the spatial distribution of edge states in Fig. 5 can be misleading at first, as it appears to be different for different but equivalent boundaries. This behavior, however, is attributed to the non-trivial temporal dynamics in the modulated non-Hermitian lattices. Thus, the evolution in time is presented not by a simple $e^{i\epsilon t}$ dynamics, but should be properly calculated by applying the non-unitary evolution operator $\widehat{U}(t)$. As a result of this non-unitary dynamics, the wave-function exhibits additional variations in time due to alternating attenuation and growth cycles, which take place at different times for different atoms of the lattice. This complex dynamics can be understood in a simple language as the result of amplification and decay when a particular atom of the lattice enters a period of gainy or lossy response, leading to a local growth or decay of the field amplitude. Direct application of the evolution operator to the instantaneous (stroboscopic) wavefunction confirms that the field profiles on different cuts are equivalent, with a phase shift of $\pm 120$ deg. (and temporal shift of T/3) between them. Moreover, the energy density averages over one cycle of gain-loss modulation, leading to a uniform field profile of the edge states (not shown). The case of temporal dynamics of the edge states with complex-valued spectrum is of special interest. The non-vanishing imaginary part of their quasi-frequency implies that the edge states exponentially grow or decay in time. Indeed, directly applying the evolution operator to complex edge states, we find that over time the energy density of the state experiences a uniform (synchronous) exponential growth or decay at all sites of the lattice.

The proposed non-Hermitian Floquet scheme can be readily implemented in a variety of systems. In particular, radio-frequency (RF) and acoustic systems with gain and loss have been of significant recent interest in the context of PT symmetric structures, and a number of successful experimental realizations have been reported. On the other hand, modulation of Hermitian RF and acoustic systems was of a separate interest, due to the possibility to achieve non-reciprocal responses and a few prototype non-reciprocal devices have been presented [33]. Combining these two ideas should be straightforward. A more challenging task, however, is to translate this concept to higher frequencies, for example aiming at optical applications. Here, the main restriction comes from the limited modulation speed of the material parameters. For graphene, the modulation of absorption with a periodically modulated pump field through saturable absorption is possible, yet it is limited by the relaxation time of carriers in graphene $\tau_r \sim 1$ ps, which sets the upper limit of modulation of a few to ten THz, and therefore the proposed scheme may be realized in the mid-IR domain. A promising path to push this idea further into near-IR and visible frequencies is to utilize optomechanically induced coherent photon–phonon gain, which has been recently used in the experimental realization of nonreciprocal amplifying responses [53, 54].

In summary, we have shown that time-modulated non-Hermitian potentials can lead to the emergence of unique topological regimes associated with the presence of exceptional points in the edge spectrum. New topological Floquet regimes have been shown to yield amplifying edge transport and lossless robust edge propagation, despite the non-Hermiticity of the lattice. These results can therefore be of immense importance for practical applications, such as for robust lasers and nonreciprocal devices.

**Data availability**

Data that are not already included in the paper and/or in the Supplementary Information are available on request from the authors.

**Author contributions**

All authors contributed extensively to the work presented in this paper.

**Acknowledgements**

The work was supported by the National Science Foundation grants DMR-1809915 and EFRI-1641069. Research carried out in part at the Center for Functional Nanomaterials, Brookhaven National Laboratory, which is supported by the U.S. Department of Energy, Office of Basic Energy Sciences, under Contract No. DE-SC0012704.

**Competing interests**
The authors declare no competing interests.

**Corresponding authors**
Correspondence to Alexander B. Khanikaev.

# Supplementary materials

## S1. Calculation of the Berry curvature and Chern numbers

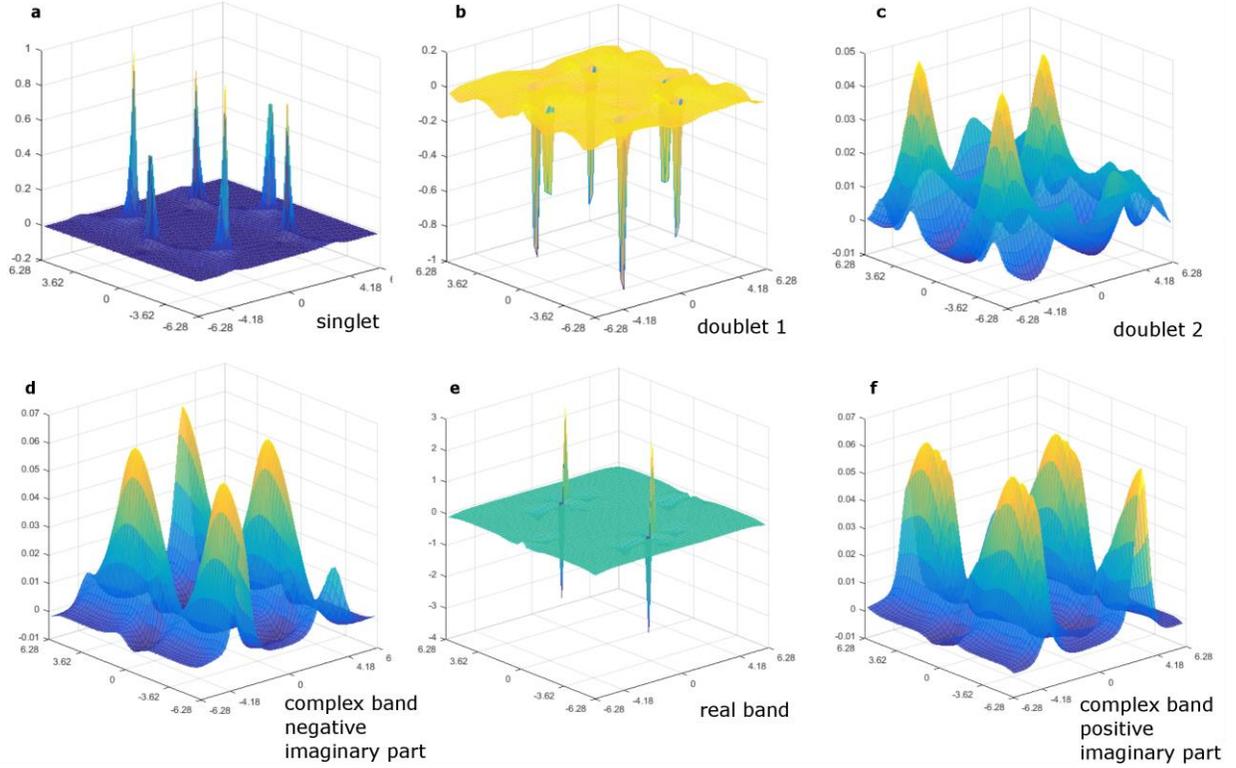

**Fig. 6 | Berry curvature for weak and strong modulation bands (temporary pic).** Chern numbers for weak modulation (**a**, **b**, **c**) and strong modulation (**d**, **e**, **f**) cases are calculated, with the same group of values $C = (1, -2, 1)$ for each case.

## S1. Derivation of the effective Hamiltonian near the $\Gamma$-point

First, to disentangle dipolar modes from the lower-frequency monopole, the unitary transformation is applied to the Hamiltonian $\widehat{\mathcal{H}}_0$ and to the time-dependent potential $\widehat{V}_S(t)$:

$$\widehat{U} = \frac{1}{\sqrt{3}} \begin{pmatrix} 1 & 1 & 1 \\ 1 & e^{-i\frac{2}{3}\pi} & e^{i\frac{2}{3}\pi} \\ 1 & e^{i\frac{2}{3}\pi} & e^{-i\frac{2}{3}\pi} \end{pmatrix}$$

The resultant 2x2 Hamiltonian acts on circularly polarized dipolar modes and, up to the second order in wavenumber, has the form

$$\widehat{\mathcal{H}}'_0(\mathbf{k}) = \begin{pmatrix} \frac{j}{4}(k_x^2 + k_y^2) & -\frac{i}{2}jk_xk_y + \frac{j}{4}(-k_x^2 + k_y^2) \\ \frac{i}{2}jk_xk_y + \frac{j}{4}(-k_x^2 + k_y^2) & \frac{j}{4}(k_x^2 + k_y^2) \end{pmatrix},$$

while the resultant reduced piecewise potential in interaction picture has the form

$$V_{I1}'(\mathbf{k}) = v \begin{pmatrix} \frac{i}{2}jk_xk_y & \frac{i}{2} \\ \frac{i}{2} & \frac{i}{2}jk_xk_y \end{pmatrix},$$

$$V_{I2}'(\mathbf{k}) = v \begin{pmatrix} \frac{i\sqrt{3}}{8}j(-k_x^2 + k_y^2) + \frac{i}{4}jk_xk_y & \frac{(-i-\sqrt{3})}{4} \\ \frac{(-i+\sqrt{3})}{4} & -\frac{i\sqrt{3}}{8}j(-k_x^2 + k_y^2) - \frac{i}{4}jk_xk_y \end{pmatrix},$$

$$V_{I3}'(\mathbf{k}) = v \begin{pmatrix} \frac{i\sqrt{3}}{8}j(k_x^2 - k_y^2) + \frac{i}{4}jk_xk_y & \frac{(-i+\sqrt{3})}{4} \\ \frac{(-i-\sqrt{3})}{4} & -\frac{i\sqrt{3}}{8}j(k_x^2 - k_y^2) - \frac{i}{4}jk_xk_y \end{pmatrix},$$

At the $\Gamma$ point, assuming weak modulation ($v \ll 1$), we find that the largest corrections are of the 2nd order in $v$, and the gain-loss induced modulation leads to the correction to the effective Hamiltonian of the form

$$\hat{V}_{eff}' = \begin{pmatrix} \frac{\sqrt{3}}{8}v^2 & 0 \\ 0 & -\frac{\sqrt{3}}{8}v^2 \end{pmatrix} + O(v^3) + \cdots,$$

which allows us to write the effective Hamiltonian can be written in Pauli basis as

$$\hat{H}_{eff}' = \frac{j(k_x^2+k_y^2)}{4}\hat{\sigma}_0 + \frac{j(k_y^2-k_x^2)}{4}\hat{\sigma}_x + \frac{jk_xk_y}{2}\hat{\sigma}_y + \frac{\sqrt{3}}{8}v^2\hat{\sigma}_z,$$

with $\hat{\sigma}_z$ term playing the role of an effective magnetic field opening topological band gap between dipolar bands at $\Gamma$ point. Importantly, the effective potential $\frac{\sqrt{3}}{8}v^2$ is a real number despite the fact that the modulation applied to the system was purely imaginary.